\documentclass[3p,twocolumn]{elsarticle}
\usepackage{natbib}
\usepackage{amsmath}
\usepackage{caption}
\usepackage{color}
\usepackage{ulem}
\usepackage{makecell}
\newcommand{\bvec}[1]{\mbox{\boldmath $#1$}}

\usepackage{amssymb}

\begin{document}

\begin{frontmatter}

\title{Choice of basis vectors for conventional unit cells revisited}

\author[myprimaryaddress]{Yoyo Hinuma}
\ead{yoyo.hinuma@gmail.com}

\author[mysecondaryaddress]{Atsushi Togo}

\author[myprimaryaddress]{Hiroyuki Hayashi}

\author[myprimaryaddress,mysecondaryaddress,mythirdaryaddress]{Isao Tanaka\corref{mycorrespondingauthor}}

\address[myprimaryaddress]{Department of Materials Science and Engineering, Kyoto University, Sakyo, Kyoto 606-8501, Japan}
\address[mysecondaryaddress]{Elements Strategy Initiative for Structure Materials (ESISM), Kyoto University, Sakyo, Kyoto 606-8501, Japan}
\address[mythirdaryaddress]{Nanostructures Research Laboratory, Japan Fine Ceramics Center, Atsuta, Nagoya 456-8587, Japan}

\begin{abstract}
Band diagrams often pass through special $\mathbf{k}$-vector points with
``irrational'' fractional coordinates such as some Brillouin zone vertices and
centers of edges. Use of ``standard'' primitive cells as defined by
Setyawan and Curtarolo [Comp. Mater. Sci. 49, 299] is convenient for
band diagram and effective mass calculations because the definitions and
coordinates of special $\mathbf{k}$-vector points including ``irrational''
fractional points are provided, but their basis vectors are taken differently
and/or special $\mathbf{k}$-vector point definitions differ from the
crystallographic convention in many cases. On the other hand, the Bilbao
Crystallographic Server [Bulg. Chem. Commun. 43, 183] defines special
$\mathbf{k}$-vector points based on crystallographic convention;
however, symbols for ``irrational'' special $\mathbf{k}$-vector points are
not defined. Obviously there is a need for a crystallographic
convention-friendly scheme to describe such ``irrational'' special
$\mathbf{k}$-vector points. Therefore, we derived a procedure to uniquely choose basis vectors for all space groups from the crystal symmetry, basis vector lengths, and interaxial angles, and implemented this algorithm in the spglib code [http://spglib.sourceforge.net/]. This paper outlines an algorithm that is
relatively straightforward to implement in software to bridge between crystallographic conventional
cells and 
``standard'' primitive cells by retaking basis
vectors. The procedure is expected to ease systematic electronic and
phonon band calculations.
\end{abstract}
\end{frontmatter}

\section{Introduction}
\label{sec:introduction}

An algorithm for efficient and systematic generation of electronic and
phonon band diagrams would be useful not only in detailed investigation
of certain compounds but also in combinatorial or ``high-throughput''
calculations \cite{NM,Hautier}. Band diagrams and effective carrier
masses are typically obtained by sampling $\mathbf{k}$-points along a path spanning the first Brillouin zone (BZ). Although the
shape of a BZ is independent on how the basis vectors are chosen, the
fractional coordinates of vertices and centers of edges and faces, hereafter
denoted as special $\mathbf{k}$-vector points, are not. Therefore, rules
to determine basis vectors that give the same special
$\mathbf{k}$-vector point coordinates for a given form of a BZ would be
useful. The choice of basis vectors, definitions of special
$\mathbf{k}$-vector points, and suggested band diagram paths are
provided in detail for ``standard'' primitive cells as defined by
Setyawan and Curtarolo [3]. However, the ``standard'' cell differs from
the crystallographic conventional cell in many cases, and for this
reason we use the term ``standard'' with quotation marks in this article
when referring to Setyawan and Curtarolo'’s definition. Consequently,
there is demand for a simple procedure, preferably using transformation
matrices for retaking basis vectors, to convert a crystallographic
conventional cell into a ``standard'' cell. We note that the AFLOW code
\cite{SC} can automatically generate ``standard'' cells; however,
details of the algorithm to derive a ``standard'' cell is not disclosed.

The objective of this article is to review how the conventional and
``standard'' definitions are different and to outline an algorithm to
convert a crystallographic conventional unit cell of the standard
setting into a ``standard'' primitive cell. This algorithm is
comprehensive and relatively easy to implement in
software. Section~\ref{sec:summary-of-definitions} provides a summary of
definitions and notations used in this paper.
Section~\ref{sec:convention} introduces the crystallographic convention
on basis vector choice. Section~\ref{sec:deriving} outlines the
differences between crystallographic and ``standard'' conventional cells
for each Bravais lattice and gives procedures and transformation
matrices to convert the former to the
latter. Section~\ref{sec:standard-basis} outlines how ``standard''
primitive basis vectors are expressed in Cartesian coordinates. The
definitions are based on Setyawan and Curtarolo \cite{SC} but re-written
in a form more convenient for implementation in
software. Section~\ref{sec:lattice-constant} shows criteria that
determine the topology of the
BZ. Section~\ref{sec:issues-regarding} brings up the necessity of
handling special $\mathbf{k}$-vector points with ``irrational''
coordinates and discusses issues arising in comparing special
$\mathbf{k}$-vector point symbols between those in the Bilbao
Crystallographic Server \cite{Bilbao2,Bilbao,CDML} and Setyawan and
Curtarolo \cite{SC}. Section~\ref{sec:summary} is a summary of this
paper.

\section{Symbols and terminologies}
\label{sec:summary-of-definitions}

Definitions of symbols are summarized in Table
\ref{table:symbol-definitions}. Symbols for basis vectors are written in
this article as $\mathbf{a}_1$, $\mathbf{a}_2$, and $\mathbf{a}_3$ in
line with the Fifth edition of the International Tables of
Crystallography A (ITA) \cite{ITA} and the Third edition of the
International Tables of Crystallography B (ITB) \cite{ITB}. The
``standard'' conventional and primitive cells are defined according to
Setyawan and Curtarolo \cite{SC}. In essence, reciprocal space basis
vectors of ``standard'' primitive cells penetrate the center of Bragg
planes belonging to the first BZ and direct space basis vector lengths
are chosen as $a'<b'<c'$ when there is ambiguity. Primes are added for
``standard'' cells and the subscript P indicates a primitive cell. All
direct space basis vectors are column vectors and reciprocal space basis
vectors are row vectors. The direct and reciprocal space basis vectors
are related by $\mathbf{a}_i^* \cdot \mathbf{a}_j = \delta_{ij}$,
$\mathbf{a}_i'^* \cdot \mathbf{a}_j' = \delta_{ij}$,
$\mathbf{a}_{\mathrm{P}i}^* \cdot \mathbf{a}_{\mathrm{P}j} =
\delta_{ij}$ ($i, j \in \{1, 2, 3\}$) using the Kronecker delta symbol
$\delta_{ij}$. Lengths of basis vectors and interaxial angles are
collectively referred to as lattice constants.

The symbols of the Bravais lattices and the sections where topics are
discussed in this article with respect to each Bravais lattice are summarized in
Table~\ref{table:bravais-symbols}.

\begin{table*}[ht]
 \caption{\label{table:symbol-definitions} List of symbols used
 in this study.}
\centering
 \begin{tabular}{lccc}
\hline 
\hline 
Definition & Basis vectors & Basis vector lengths & Interaxial angles \\
  \hline 
  Crystallographic conventional cell & $\{\mathbf{a}_1, \mathbf{a}_2,
      \mathbf{a}_3\}$ & $\{a, b, c\}$ & $\{\alpha, \beta, \gamma\}$ \\
  Reciprocal crystallographic conventional cell & $\{\mathbf{a}^*_1,
      \mathbf{a}^*_2, 
      \mathbf{a}^*_3\}$ & $\{k_a, k_b, k_c\}$ & $\{k_\alpha, k_\beta,
              k_\gamma\}$ \\
  ``standard'' conventional cell & $\{\mathbf{a}'_1, \mathbf{a}'_2,
      \mathbf{a}'_3\}$ &
          $\{a', b', c'\}$ & $\{\alpha', \beta', \gamma'\}$ \\
  Reciprocal ``standard'' conventional cell & $\{\mathbf{a}'^*_1, \mathbf{a}'^*_2,
      \mathbf{a}'^*_3\}$ &
          $\{k'_a, k'_b, k'_c\}$ & $\{k'_\alpha, k'_\beta, k'_\gamma\}$ \\
  ``standard'' primitive cell & $\{\mathbf{a}'_{\mathrm{P}1},
      \mathbf{a}'_{\mathrm{P}2}, 
      \mathbf{a}'_{\mathrm{P}3}\}$ &
          $\{a'_\mathrm{P}, b'_\mathrm{P}, c'_\mathrm{P}\}$ & $\{\alpha', \beta',
              \gamma'\}$ \\
  Reciprocal ``standard'' primitive cell & $\{\mathbf{a}'^*_{\mathrm{P}1},
      \mathbf{a}'^*_{\mathrm{P}2}, \mathbf{a}'^*_{\mathrm{P}3}\}$ &
          $\{k'_{\mathrm{P}a}, k'_{\mathrm{P}b}, k'_{\mathrm{P}c}\}$ &
              $\{k'_{\mathrm{P}\alpha}, k'_{\mathrm{P}\beta},
              k'_{\mathrm{P\gamma}}\}$  \\
\hline 
\hline
 \end{tabular}
\end{table*}

\begin{table*}[ht]
 \caption{\label{table:bravais-symbols} Symbols of the
 Bravais lattices and sections of this article where they are discussed.}
 \centering
 \begin{tabular}{llccl}
  \hline 
  \hline 
  Crystal system & Centring & Symbol && Sections \\
  \hline 
  Triclinic & Primitive & $aP$ && \ref{sec:convention-triclinic}, \ref{sec:deriving-aP}, \ref{sec:lattice-constant-aP} \\
  Monoclinic & Primitive & $mP$ && \ref{sec:convention-monoclinic}, \ref{sec:deriving-mP}, \ref{sec:lattice-constant-P} \\ 
  & Base-centred & $mS$ && \ref{sec:convention-monoclinic}, \ref{sec:deriving-mS}, \ref{sec:standard-prim-mS}, \ref{sec:lattice-constant-mS} \\
  Orthorhombic & Primitive & $oP$ && \ref{sec:convention-orthorhombic}, \ref{sec:deriving-o}, \ref{sec:lattice-constant-P} \\
  & Base-centred & $oS$ && \ref{sec:convention-orthorhombic}, \ref{sec:deriving-oS}, \ref{sec:standard-prim-oS}, \ref{sec:lattice-constant-P}\\
  & Body-centred & $oI$ && \ref{sec:convention-orthorhombic}, \ref{sec:deriving-o}, \ref{sec:standard-prim-I}, \ref{sec:lattice-constant-P} \\
  & Face-centred & $oF$ && \ref{sec:convention-orthorhombic}, \ref{sec:deriving-o}, \ref{sec:standard-prim-F}, \ref{sec:lattice-constant-oF} \\
  Tetragonal & Primitive & $tP$ && \ref{sec:convention-others}, \ref{sec:deriving-same}, \ref{sec:lattice-constant-P} \\
  & Body-centred & $tI$ && \ref{sec:convention-others}, \ref{sec:deriving-same}, \ref{sec:standard-prim-I}, \ref{sec:lattice-constant-tI} \\
  Hexagonal & Primitive & $hP$ && \ref{sec:convention-others}, \ref{sec:deriving-same}, \ref{sec:lattice-constant-P} \\
  Rhombohedral & * & $hR$ && \ref{sec:convention-others}, \ref{sec:deriving-hR}, \ref{sec:lattice-constant-hR} \\
  Cubic & Primitive & $cP$ && \ref{sec:convention-others}, \ref{sec:deriving-same}, \ref{sec:lattice-constant-P} \\
  & Body-centred & $cI$ && \ref{sec:convention-others}, \ref{sec:deriving-same}, \ref{sec:standard-prim-I}, \ref{sec:lattice-constant-P} \\
  & Face-centred & $cF$ && \ref{sec:convention-others}, \ref{sec:deriving-same}, \ref{sec:standard-prim-F}, \ref{sec:lattice-constant-P} \\
  \hline 
  \hline
  *Primitive or triple hexagonal.
 \end{tabular}
\end{table*}

A change of basis matrix $\bvec{M}$ relates a choice of basis
vectors to another choice of basis vectors. $\bvec{M}$ is a $3\times
3$ matrix and is used as
\begin{align}
(\mathbf{a}'_1, \mathbf{a}'_2, \mathbf{a}'_3) = (\mathbf{a}_1, \mathbf{a}_2, \mathbf{a}_3) \bvec{M}.
\end{align}
A transformation matrix, $\bvec{P}$, transforms a conventional unit cell
to one of the primitive cells by
\begin{equation}
(\mathbf{a}'_{\mathrm{P}1},
\mathbf{a}'_{\mathrm{P}2}, \mathbf{a}'_{\mathrm{P}3}) = 
(\mathbf{a}'_1, \mathbf{a}'_2, \mathbf{a}'_3) \bvec{P}.
\end{equation}
With the exception of rhombohedral cells, the change of basis $\bvec{M}$
conserves the number of atoms in the unit cell, $n_\mathrm{a}$, because
$\det(\bvec{M})=1$. On the other hand, $\det(\bvec{P})=1$ for cells without
centring, $\det(\bvec{P})=1/2$ for base- and body-centred cells and
$\det(\bvec{P})=1/4$ for face-centred cells. The number of atoms in the
primitive cell is $n_\mathrm{a}\det(\bvec{P})$.

The unit cell is a parallelepiped determined by the basis vectors
\cite{ITA}. A point in the unit cell $\bvec{x}$ is represented
by a column vector of fractional values with respect to basis vector
lengths:
\begin{equation}
\label{eq:atom-point}
\bvec{x}^{\mathrm{T}}=(x_1, x_2, x_3)\;\;\text{where}\;\; 0 \leq x_i < 1.
\end{equation}The position of an atom at a point
$\bvec{x}$ in the unit cell is given by
\begin{equation}
\label{eq:atom-position}
 \bvec{X} = \sum_i x_i \mathbf{a}_i = (\mathbf{a}_1, \mathbf{a}_2, \mathbf{a}_3)\bvec{x}.
\end{equation}
$\bvec{X}$ is invariant with change of basis. However, as
\begin{align}
(\mathbf{a}_1, \mathbf{a}_2, \mathbf{a}_3) \bvec{x}  = 
(\mathbf{a}'_1, \mathbf{a}'_2, \mathbf{a}'_3) \bvec{M}^{-1} \bvec{x},
\end{align}
the vector representation of the atom varies with change of basis
$\bvec{M}$ according to
\begin{equation}
 \label{eq:atom-point-transformation}
  \bvec{x}' = \bvec{M}^{-1} \bvec{x} + \bvec{t},
\end{equation} 
where $\bvec{t}$ is the lattice translation vector necessary to fulfill the
condition (\ref{eq:atom-point}) so that the point is in
the new unit cell.

Irrational special $\mathbf{k}$-vector points are defined as
 special $\mathbf{k}$-vector points that always
 have irrational fractional coordinates
irrespective of basis vector lengths except in virtual crystals. Some vertices of the BZ are
located at irrational special $\mathbf{k}$-vector
points in body-centred tetragonal, face-, body-, and base-centred
orthorhombic, rhombohedral, monoclinic, base-centred monoclinic, and
triclinic Bravais lattices.

\section{Convention on basis vector choice}
\label{sec:convention}

There is a crystallographic convention on which specific
choice of basis vectors should be used out of an infinite number of
choices, and this is outlined in the ITA \cite{ITA} and ITB \cite{ITB}.
Here, the Bravais lattice is determined, symmetry operations are
identified, and then the origin is decided. A total of 530 choices,
distinguished by Hermann-Mauguin and Hall symbols \cite{ITB,Hall}, of
230 space group types are listed in the ITA and ITB. Multiple choices
are provided for many space group types. For instance, space group type
$Pnn2$ (No. 34) can be $Pn2n$ or $P2nn$ depending on how the basis
vectors are chosen, and space group type $P4/ncc$ (No. 130) has two
choices with different origins. In general, changing the basis may
change the centring in base-centred cells but does not in body- or
face-centred cells; one exception is that body-centred monoclinic cells
can be converted into base-centred cells by using a different
choice. The standard choice for monoclinic cells always has unique axis
$b$ and $\beta>90^\circ$ and base-centred cells are always
$C$-centred. Base-centred orthorhombic cells are $C$-centred with the
exception of four space group numbers 38 to 41; the conventional setting
of these four space group types is $A$-centred such that the point group
is $mm2$ and the proper two-fold rotation axis is along the
$c$-axis. The triple hexagonal unit cell for rhombohedral
 Bravais lattices could be obverse or
reverse; the current version of the International Tables takes the
obverse setting as the standard setting although the reverse setting was
the standard until 1952 \cite{ITA}.

The ITA by itself does not 
uniquely define the crystallographic conventional cell in a number of
situations. For instance, the order of basis vector lengths is not
specified, which becomes problematic in many monoclinic and orthorhombic
space group types. One standardization algorithm to uniquely determine
the conventional cell is proposed by Parth\'{e} \textit{et
al}. \cite{PartheJAC} They require the cell to be right-handed and have
the following constraints: the Niggli reduced cell is used for triclinic
structures, ``best'' cell with unique axis $b$ and $\beta>90^\circ$
\cite{PartheBest} is used for monoclinic structures, that is, the cell
where $a$ and $c$ are as short as possible while conforming to the
standard space group symbol, $a \leq b$ or $a \leq b \leq c$ for
orthorhombic structures, and obverse triple hexagonal cell for
rhombohedral structures. Additionally, origin 2 is used for
centrosymmetric structures. The space group type with the smallest index
for the relevant screw axis is taken. For instance, one can choose to
define space group number 80 as $I4_{3}$ but, as this space group type
can also be defined as $I4_{1}$, the former is not accepted as a choice
of the space group type and the latter is adopted because the screw axis
index is smaller. Parth\'{e} \textit{et al}. \cite{PartheJAC} provides
further rules to reduce the values of internal coordinates that we will
not discuss in this paper. The above procedure unambiguously defines the
basis vectors of the crystallographic conventional cell but the internal
coordinates are not necessarily determined. One example is CsCl with
space group type $Pm\bar{3}m$ as mentioned in Parth\'{e} and Gelato
\cite{PartheStandard}. Wyckoff positions $1a$ and $1b$, which both have
the same site symmetry $m3m$, must be occupied by Cs or Cl, but the
above requirement does not determine whether Cs should occupy $1a$ or
$1b$ sites. This ambiguity still remains after the standardization
procedure of Parth\'{e} \textit{et al}. \cite{PartheJAC} However, the band
diagram is not affected by
the choice of origin since this is a physical property
in reciprocal space that is invariant under translation in direct space.
The conventional cells employed in this work conform to the conventional
coordinate system in Table 2.1.2.1 of the ITA \cite{ITA}. The basis
vectors cannot be uniquely defined from Table 2.1.2.1 in many space
group types, and therefore the following additional rules are
imposed. We take the first choice that appears in Table A1.4.2.7 of the
ITB \cite{ITB} as the standard choice. This means that origin 1 is
always used and the space group type with the smallest index for the
relevant screw axis is automatically taken. All other choices in a
particular space group type can be converted into the first choice by
change of basis and/or translation of the origin. Table 9.3.7.1 of the
ITA provides definitions of conventional cells based on lattice
constants; however, the resulting space group choice is not necessarily
the first one in Table A1.4.2.7 of the ITB \cite{ITB}, therefore the
definition in Table 9.3.7.1 of the ITA is not adopted. Parth\'{e}
\textit{et al}. \cite{PartheJAC} instead takes the symmetry center at
the origin, that is, adopt origin choice 2, for centrosymmetric
structures. Origin choice 2 is never the first choice that appears in
Table A1.4.2.7 of the ITB \cite{ITB}. Parth\'{e} \textit{et
al}. \cite{PartheJAC} also requires that ``orthorhombic structures where
the space group symbols do not prescribe a particular labelling of any
of the three unit-cell axes have to be described with a unit cell where
$a < b < c$''; however, this condition is inappropriate in some
orthorhombic space group types as discussed below.
Conventional cells are defined as above in the 
spglib code \cite{spglib-project} versions 1.7.3 and higher.

\subsection{Orthorhombic lattices}
\label{sec:convention-orthorhombic}

Some space group types have ambiguity on how to take basis vectors. An
obvious example is space group type $P222$ (No. 16)
where the six permutations of $a$, $b$, and $c$ do not change the space
group type.

The conditions to resolve ambiguity in basis vector lengths of orthorhombic space groups, with or without centring, other than $Pbca$ (No. 61) and
$Ibca$ (No. 73)
are $a<b$, and if necessary, additionally $b<c$. For example, only the former is necessary in $Pcc2$
(No. 27) and both is necessary in $Fmmm$ (No. 69). The imposed condition
for $Pbca$ and $Ibca$ is to make $a$ the shortest. Two choices for
$Pbca$ exist: $Pbca$ (adopted choice) and $Pcab$. If $Pbca$ is $a<b<c$
then $Pcab$ is $a<c<b$, and if $Pcab$ is $a<b<c$ then $Pbca$ is
$a<c<b$. Therefore, $a<b<c$ cannot be enforced if $Pbca$ is to be used
as the adopted choice. Similar logic applies to $Ibca$ (No. 73). The
point groups that are allowed for an orthorhombic lattice are $222$,
$mm2$, and $mmm$. As the
proper two-fold rotation axis is taken along the $c$-axis in the point group $mm2$, $Amm2$,
$Abm2$, $Ama2$, and $Aba2$ (numbers 38 to 41) become $A$-centred.

\subsection{Monoclinic lattices}
\label{sec:convention-monoclinic}

The definition according to Parth\'{e} and Gelato \cite{PartheBest} is
adopted and is summarized as follows. The $b$-axis is taken as the unique axis
 and $\beta>90^\circ$, and the basis vectors corresponding to
$a$ and $c$ are taken as short as possible. Base centring, if it exists, is always
$C$-centred. The condition $a<c$ is imposed if the $c$-glide symmetry
element identifiable by the ``$c$'' in the Hermann-Mauguin symbol is absent.

\subsection{Triclinic lattices}
\label{sec:convention-triclinic}

The crystallographic conventional cell for a triclinic system is the
Niggli reduced cell that can be robustly obtained using the algorithm by
K\v{r}iv\'{y} and Gruber \cite{Krivy} and its modified version that
addresses numerical errors arising in finite precision floating point
algebra by Grosse-Kunstleve \textit{et al}. \cite{GK} Parth\'{e}
\textit{et al}. \cite{PartheJAC} also employs the Niggli-reduced cell.

\subsection{Cubic, tetragonal, hexagonal, and rhombohedral lattices}
\label{sec:convention-others}

For cubic, tetragonal, and hexagonal lattices, the usual
conditions are adopted when choosing basis vectors in this article, that are,
\begin{align*}
 a=b=c \;\;\text{and}\;\;\alpha=\beta=\gamma=90^\circ\;\;&\text{for
 cubic}, \\
 a=b
 \;\;\text{and}\;\;\alpha=\beta=\gamma=90^\circ\;\;&\text{for
 tetragonal}, \\
 a=b, \alpha=\beta=90^\circ
 \;\;\text{and}\;\;\gamma=120^\circ \;\;&\text{for hexagonal},
\end{align*}
respectively. For primitive rhombohedral cells, the condition is
\begin{equation}
  a=b=c\;\;\text{and}\;\;\alpha=\beta=\gamma. \nonumber
\end{equation}
However, the conventional rhombohedral cell is defined on a hexagonal lattice. The
obverse setting of the conventional rhombohedral cell can be recovered
from the primitive rhombohedral cell using the transformation matrix
$\bvec{M}^{-1}$ of Eq.~(\ref{eq:primitive-rhombohedral}) shown in
Sec.~\ref{sec:deriving-hR}.

\section{Deriving basis vectors of the ``standard'' cell}
\label{sec:deriving}

The crystallographic conventional unit cell as defined in
Secs.~\ref{sec:convention-orthorhombic} to \ref{sec:convention-others}
is chosen as the starting point except in triclinic cells as discussed in section \ref{sec:deriving-aP}. Computer programs typically represent basis vectors in Cartesian
coordinates and store internal coordinates of atoms using
fractional coordinates or Cartesian coordinates. In this case, converting between
fractional and Cartesian coordinates is easily done by
 a linear transformation
such as Eq.~(\ref{eq:atom-position}). Standardization of basis vectors
is carried out by choosing different basis vectors, and the positions of
atoms are invariant in Cartesian coordinates during retaking of basis
vectors.  The volume of the unit cell is kept
constant with the exception of rhombohedral cells.

We will not discuss cases when the basis vectors cannot be uniquely
defined. This can happen in virtually designed crystals; one example is
an orthorhombic cell with space group type $I222$ and $a=b=c$. The basis
vectors cannot be uniquely defined unless $a$, $b$, and $c$ are all
different from each other in this case.

The first step of the outlined algorithm is identification of the
Bravais lattice, which can be deduced easily from the
space group number. The cell is triclinic
if the space group number is 1 or 2, monoclinic if 3 to 15, orthorhombic
if 16 to 74, tetragonal if 75 to 142, hexagonal or rhombohedral if 143
to 194, and cubic otherwise. The first character of the Hermann-Mauguin
symbol gives the centring ($P$, $A$, $C$, $I$, $F$, or $R$) and
distinguishes hexagonal and rhombohedral cells for space group numbers
143 to 194.

\subsection{$cP$, $cF$, $cI$, $tP$, $tI$, and $hP$}
\label{sec:deriving-same}

The crystallographic and ``standard'' \cite{SC} conventional cells are
the same for these Bravais lattices; therefore, the change of basis
matrix $\bvec{M}$ is the identity matrix.

\subsection{$hR$}
\label{sec:deriving-hR}

The rhombohedral crystallographic conventional cell is obverse triple
hexagonal while the ``standard'' conventional cell is primitive
rhombohedral\cite{SC}. The change of basis matrix and the inverse are

\begin{equation}
\label{eq:primitive-rhombohedral}
\bvec{M} = 
   \begin{pmatrix}
    \frac{2}{3} & \frac{\bar{1}}{3} & \frac{\bar{1}}{3} \\
    \frac{1}{3} & \frac{1}{3} & \frac{\bar{2}}{3} \\
    \frac{1}{3} & \frac{1}{3} & \frac{1}{3}
  \end{pmatrix}, 
\bvec{M}^{-1} =
  \begin{pmatrix}
    1 & 0 & 1 \\
    \bar{1} & 1 & 1 \\
    0 & \bar{1} & 1
  \end{pmatrix},
\end{equation}
respectively. The determinant of $\bvec{M}$ is $1/3$,
which means that the ``standard'' conventional cell volume is one-third
of the crystallographic conventional cell. One simple procedure to
remove duplicate atoms in the ``standard'' cell is
 to
apply Eq.~(\ref{eq:atom-point-transformation}) and remove duplicate atoms.
\subsection{$oP$, $oF$, and $oI$}
\label{sec:deriving-o}

The order of basis vector lengths is not
 always $a<b<c$ in a
non-base centred crystallographic conventional cell whereas the order in a
``standard'' conventional cell is always $a' < b'
< c'$ \cite{SC}. The change of basis matrix $\bvec{M}$ and the inverse
are given in Table~\ref{table:symbol-ort}.

\begin{table}[ht]
 \caption{\label{table:symbol-ort}Change of basis matrix $\bvec{M}$ for non-base centred orthorhombic cells.}
\centering
 \begin{tabular}{ccc}
\hline 
\hline 
Condition & $\bvec{M}$ & $\bvec{M}^{-1}$ \\
\hline 
  $a < b < c$ &   
  $\begin{pmatrix}
    1 & 0 & 0 \\
    0 & 1 & 0 \\
    0 & 0 & 1
  \end{pmatrix}$ & 
   $\begin{pmatrix}
    1 & 0 & 0 \\
    0 & 1 & 0 \\
    0 & 0 & 1
  \end{pmatrix}$ \\
    $a < c < b$ &   
  $\begin{pmatrix}
    \bar{1} & 0 & 0 \\
    0 & 0 & 1 \\
    0 & 1 & 0
  \end{pmatrix}$ & 
   $\begin{pmatrix}
    \bar{1} & 0 & 0 \\
    0 & 0 & 1 \\
    0 & 1 & 0
  \end{pmatrix}$ \\
    $b < a < c$ &   
  $\begin{pmatrix}
    0 & 1 & 0 \\
    1 & 0 & 0 \\
    0 & 0 & \bar{1}
  \end{pmatrix}$ & 
   $\begin{pmatrix}
    0 & 1 & 0 \\
    1 & 0 & 0 \\
    0 & 0 & \bar{1}
  \end{pmatrix}$ \\
    $b < c < a$ &   
  $\begin{pmatrix}
    0 & 0 & 1 \\
    1 & 0 & 0 \\
    0 & 1 & 0
  \end{pmatrix}$ & 
   $\begin{pmatrix}
    0 & 1 & 0 \\
    0 & 0 & 1 \\
    1 & 0 & 0
  \end{pmatrix}$ \\
      $c < a < b$ &   
  $\begin{pmatrix}
    0 & 1 & 0 \\
    0 & 0 & 1 \\
    1 & 0 & 0
  \end{pmatrix}$ & 
   $\begin{pmatrix}
    0 & 0 & 1 \\
    1 & 0 & 0 \\
    0 & 1 & 0
  \end{pmatrix}$ \\
     $c < b < a$ &   
  $\begin{pmatrix}
    0 & 0 & 1 \\
    0 & \bar{1} & 0 \\
    1 & 0 & 0
  \end{pmatrix}$ & 
   $\begin{pmatrix}
    0 & 0 & 1 \\
    0 & \bar{1} & 0 \\
    1 & 0 & 0
  \end{pmatrix}$ \\
\hline 
\hline
 \end{tabular}
\end{table}

\subsection{$oS$}
\label{sec:deriving-oS}

A base-centred orthorhombic crystallographic conventional cell is
generally $C$-centred but some space group types with point group type
$mm2$, namely space group 
types $Amm2$, $Abm2$, $Ama2$, and $Aba2$ (38 to 41) are $A$-centred. The
order of basis vector lengths is not unique in some space group
types. In contrast, a ``standard'' conventional cell is always
$C$-centred and $a' < b'$ regardless of symmetry \cite{SC}. The change
of basis matrix $\bvec{M}$ and the inverse are given in Table
\ref{table:symbol-baseort}.

\begin{table}[ht]
 \caption{\label{table:symbol-baseort}Change of basis matrix $\bvec{M}$ for base centred orthorhombic cells.}
\centering
 \begin{tabular}{ccc}
\hline 
\hline 
Condition & $\bvec{M}$ & $\bvec{M}^{-1}$ \\
\hline 
  $C$-centred, $a< b$ &   
      $\begin{pmatrix}
	1 & 0 & 0 \\
	0 & 1 & 0 \\
	0 & 0 & 1
       \end{pmatrix}$ & 
  $\begin{pmatrix}
    1 & 0 & 0 \\
    0 & 1 & 0 \\
    0 & 0 & 1
   \end{pmatrix}$ \\
  $C$-centred, $b < a$ &   
      $\begin{pmatrix}
	0 & 1 & 0 \\
	1 & 0 & 0 \\
	0 & 0 & \bar{1}
       \end{pmatrix}$ & 
  $\begin{pmatrix}
    0 & 1 & 0 \\
    1 & 0 & 0 \\
    0 & 0 & \bar{1}
   \end{pmatrix}$ \\
  $A$-centred, $b < c$ &   
      $\begin{pmatrix}
	0 & 0 & 1 \\
	1 & 0 & 0 \\
	0 & 1 & 0
       \end{pmatrix}$ & 
  $\begin{pmatrix}
    0 & 1 & 0 \\
    0 & 0 & 1 \\
    1 & 0 & 0
   \end{pmatrix}$ \\
  $A$-centred, $c < b$  &   
      $\begin{pmatrix}
	0 & 0 & \bar{1} \\
	0 & 1 & 0 \\
	1 & 0 & 0
       \end{pmatrix}$ & 
  $\begin{pmatrix}
    0 & 0 & 1 \\
    0 & 1 & 0 \\
    \bar{1} & 0 & 0
   \end{pmatrix}$ \\
   \hline 
\hline
 \end{tabular}
\end{table}

\subsection{$mP$}
\label{sec:deriving-mP}

As outlined in Parth\'{e} and Gelato \cite{PartheBest}, a crystallographic
conventional cell has unique axis $b$ and $\beta > 90^\circ$. The
additional constraint of $a < c$ is always imposed on space group types
without c-glide symmetry. The order of $a$ and $c$ is not always the
same when there is $c$-glide symmetry because this symmetry element
distinguishes $\mathbf{a}_1$ and $\mathbf{a}_3$. On the other hand, a ``standard''
conventional cell has unique axis $a$, lattice constants $b' \leq
c'$, and $\alpha' < 90^\circ$ \cite{SC}. The change of basis matrix
$\bvec{M}$ and the inverse are given in  Table \ref{table:symbol-mono}.

\begin{table}[ht]
 \caption{\label{table:symbol-mono}Change of basis matrix $\bvec{M}$ for monoclinic cells.}
\centering
 \begin{tabular}{ccc}
\hline 
\hline 
Condition & $\bvec{M}$ & $\bvec{M}^{-1}$ \\
\hline 
  $a < c$ &   
  $\begin{pmatrix}
   0 & \bar{1} & 0 \\
    1 & 0 & 0 \\
    0 & 0 & 1
  \end{pmatrix}$ & 
   $\begin{pmatrix}
    0 & 1 & 0 \\
    \bar{1} & 0 & 0 \\
    0 & 0 & 1
  \end{pmatrix}$ \\
    $c< a$  &   
  $\begin{pmatrix}
    0 & 0 & \bar{1} \\
    \bar{1} & 0 & 0 \\
    0 & 1 & 0
  \end{pmatrix}$ & 
   $\begin{pmatrix}
    0 & \bar{1} & 0 \\
    0 & 0 & 1 \\
    \bar{1} & 0 & 0
  \end{pmatrix}$ \\
  
   \hline 
\hline
 \end{tabular}
\end{table}

\subsection{$mS$}
\label{sec:deriving-mS}

A crystallographic conventional cell is $C$-centred with unique axis $b$
and $\beta > 90^\circ$, which is required in a ``best'' cell as defined by
Parth\'{e} and Gelato \cite{PartheBest}.  In contrast, a ``standard''
conventional cell is $C$-centred with unique axis $a$ and $\alpha' <
90^\circ$ \cite{SC}. We believe the condition $b' \leq c'$ in the
original paper by Setyawan and Curtarolo cannot be enforced because the
basis vector length of $\mathbf{a}'_2$ in the plane with centring is not
necessarily shorter than the basis vector length of $\mathbf{a}'_3$
without centring. The three basis vectors are distinguished by symmetry;
one of the basis vectors comprising the plane with centring forms a
right angle with the unique axis while the other does not. The change of
basis matrix and its inverse are

\begin{equation}
\bvec{M} = 
   \begin{pmatrix}
    0 & \bar{1} & 0 \\
    1 & 0 & 0 \\
    0 & 0 & 1
  \end{pmatrix}, 
\bvec{M}^{-1} =
  \begin{pmatrix}
    0 & 1 & 0 \\
    \bar{1} & 0 & 0 \\
    0 & 0 & 1
  \end{pmatrix}.
\end{equation}
The reason why only one change of basis
matrix is sufficient is provided in \ref{sec:appendix-a}.

\subsection{$aP$}
\label{sec:deriving-aP}

The crystallographic conventional triclinic cell is the Niggli reduced
cell \cite{PartheJAC,Krivy, GK} that is determined with direct space
lattice constants. The order of basis vector lengths are $a < b < c$,
and $\alpha$, $\beta$, and $\gamma$ are all smaller than (Type I) or
larger than (Type II) $90^\circ$. On the other hand, the ``standard''
conventional cell is defined based on reciprocal basis vectors. The
requirement that reciprocal basis vectors must penetrate centers of faces of
the first BZ means that the reciprocal
``standard'' conventional cell must be a Niggli reduced cell in
reciprocal space with axes relabeled. In other words, the set of basis
vector lengths of the reciprocal ``standard'' conventional cell must
match that of the Niggli reduced cell in reciprocal space. Moreover, the
interaxial angles of the reciprocal ``standard'' conventional cell,
$k'_\alpha$, $k'_\beta$, and $k'_\gamma$, must be either all larger than
or smaller than $90^\circ$ and $k'_\gamma$ is always the one closest to
$90^\circ$ \cite{SC}.

The conversion from crystallographic to ``standard'' cells is done in
the following three steps. First, the Niggli reduced cell in reciprocal
space is obtained and its reciprocal basis vectors, that is, basis
vectors in direct space, are used as $\mathbf{a}_1$, $\mathbf{a}_2$, and
$\mathbf{a}_3$ instead of the basis vectors of the crystallographic
conventional cell. Next, basis vectors are relabeled such that $k_\gamma$
is closest to $90^\circ$ by using a change of basis matrix $\bvec{M}_1$
that transforms to an intermediate form: $(\mathbf{a}''_1,
\mathbf{a}''_2, \mathbf{a}''_3) = (\mathbf{a}_1, \mathbf{a}_2,
\mathbf{a}_3)\bvec{M}_1$ . Finally, another change of basis matrix
$\bvec{M}_2$ is used as $(\mathbf{a}'_1, \mathbf{a}'_2, \mathbf{a}'_3)
= (\mathbf{a}''_1, \mathbf{a}''_2, \mathbf{a}''_3)\bvec{M}_2$ to
invert basis vectors as necessary such that $k_\alpha$, $k_\beta$, and
$k_\gamma$ become all larger than or smaller than $90^\circ$. In other words,
\begin{align}
(\mathbf{a}'_1, \mathbf{a}'_2, \mathbf{a}'_3) = (\mathbf{a}_1, \mathbf{a}_2, \mathbf{a}_3) \bvec{M}_1\bvec{M}_2.
\end{align}
The change of basis matrix for the second step, $\bvec{M}_1$, is given
in Table \ref{table:symbol-trimat1}. The change of basis matrix for the
third step, $\bvec{M}_2$, depends on the interaxial angles of the
reciprocal unit cell of the intermediate form $k''_\alpha$, $k''_\beta$,
and $k''_\gamma$ and is given in Table \ref{table:symbol-trimat2}.

\begin{table}[ht]
 \caption{\label{table:symbol-trimat1}Change of basis matrix $\bvec{M}_1$ for triclinic cells.}
\centering
 \begin{tabular}{ccc}
\hline 
\hline 
Condition & $\bvec{M}_1$ & $\bvec{M}_1^{-1}$ \\
\hline 
  $k'_\alpha$ closest to  $90^\circ$ &   
  $\begin{pmatrix}
   0 & 0 & 1 \\
    1 & 0 & 0 \\
    0 & 1 & 0
  \end{pmatrix}$ & 
   $\begin{pmatrix}
    0 & 1 & 0 \\
    0 & 0 & 1 \\
    1 & 0 & 0
  \end{pmatrix}$ \\
    $k'_\beta$ closest to  $90^\circ$   &   
  $\begin{pmatrix}
    0 & 1 & 0 \\
    0 & 0 & 1 \\
    1 & 0 & 0
  \end{pmatrix}$ & 
   $\begin{pmatrix}
    0 & 0 & 1 \\
    1 & 0 & 0 \\
    0 & 1 & 0
  \end{pmatrix}$ \\
    $k'_\gamma$ closest to  $90^\circ$ &   
  $\begin{pmatrix}
    1 & 0 & 0 \\
    0 & 1 & 0 \\
    0 & 0 & 1
  \end{pmatrix}$ & 
   $\begin{pmatrix}
    1 & 0 & 0 \\
    0 & 1 & 0 \\
    0 & 0 & 1
  \end{pmatrix}$ \\
  
   \hline 
\hline
 \end{tabular}
\end{table}

\begin{table}[ht]
 \caption{\label{table:symbol-trimat2}Change of basis matrix $\bvec{M}_2$ for triclinic cells.}
\centering
 \begin{tabular}{cc}
\hline 
\hline 
Condition & $\bvec{M}_2=\bvec{M}_2^{-1}$ \\
\hline 
  \makecell[l]{$k''_\alpha < 90^\circ$, $k''_\beta < 90^\circ$,
  $k''_\gamma < 90^\circ$ 
  or \\
  $k''_\alpha > 90^\circ$, $k''_\beta > 90^\circ$, $ k''_\gamma > 90^\circ$}
   &   
  $\begin{pmatrix}
    1 & 0 & 0 \\
    0 & 1 & 0 \\
    0 & 0 & 1
  \end{pmatrix}$ \\
  \makecell[l]{$k''_\alpha < 90^\circ, k''_\beta > 90^\circ, k''_\gamma > 90^\circ$ 
  or \\
  $k''_\alpha > 90^\circ, k''_\beta < 90^\circ, k''_\gamma < 90^\circ$} 
   &   
  $\begin{pmatrix}
    1 & 0 & 0 \\
    0 & \bar{1} & 0 \\
    0 & 0 & \bar{1}
    \end{pmatrix}$ \\
  \makecell[l]{$k''_\alpha > 90^\circ, k''_\beta < 90^\circ, k''_\gamma > 90^\circ$ or \\
  $k''_\alpha < 90^\circ, k''_\beta > 90^\circ, k''_\gamma < 90^\circ$} 
   &   
  $\begin{pmatrix}
    \bar{1} & 0 & 0 \\
    0 & 1 & 0 \\
    0 & 0 & \bar{1}
    \end{pmatrix}$ \\
  \makecell[l]{$k''_\alpha > 90^\circ, k''_\beta > 90^\circ, k''_\gamma < 90^\circ$ or \\ 
  $k''_\alpha < 90^\circ, k''_\beta < 90^\circ, k''_\gamma > 90^\circ$} 
       &   
  $\begin{pmatrix}
    \bar{1} & 0 & 0 \\
    0 & \bar{1} & 0 \\
    0 & 0 & 1
    \end{pmatrix}$ \\
   \hline 
\hline
 \end{tabular}
\end{table}

\section{''Standard'' basis vectors in Cartesian coordinates}
\label{sec:standard-basis}

``Standard'' basis vectors can be given in Cartesian coordinates once
the ``standard'' lattice constants are
obtained.  Representation of points in a unit cell as given in
Eq.~(\ref{eq:atom-point}) is invariant against rigid rotation of
the corresponding crystal. The following are how
``standard'' basis vectors, both conventional and primitive, are
represented in Cartesian coordinates according to the definition by
Setyawan and Curtarolo~\cite{SC}. 

\subsection{''Standard'' conventional cell}

A representation in Cartesian coordinates of the ``standard'' conventional cell, excluding
hexagonal (\textit{hP}) and rhombohedral (\textit{hR}) cells, is
\begin{align*}
 {\mathbf{a}'_1}^{\mathrm{T}} &= (a',0,0), \\
 {\mathbf{a}'_2}^{\mathrm{T}} &= (b'\cos\gamma',b'\sin\gamma',0), \\
 {\mathbf{a}'_3}^{\mathrm{T}} &= (c'_x,c'_y,c'_z),
\end{align*}
where
\begin{align*}
c'_x &= c'\cos\beta, \\
c'_y &= c'\frac{(\cos\alpha'-\cos\beta'\cos\gamma')}{\sin\gamma'}, \\
c'_z &= \\
c'&\frac{\sqrt{\sin^2\gamma'-\cos^2\alpha' - \cos^2\beta'+2\cos\alpha'\cos\beta'\cos\gamma'}}{\sin\gamma'}.
\end{align*}
For hexagonal cells, the representation is
\begin{align*}
 {\mathbf{a}'_1}^{\mathrm{T}} &= \left(\frac{a'}{2},-\frac{\sqrt{3}a'}{2},0\right), \\
 {\mathbf{a}'_2}^{\mathrm{T}} &= \left(\frac{a'}{2},\frac{\sqrt{3}a'}{2},0\right), \\
 {\mathbf{a}'_3}^{\mathrm{T}} &= (0,0,c'),
\end{align*}
and for rhombohedral cells,
\begin{align*}
 {\mathbf{a}'_1}^{\mathrm{T}} &= \left(a'\cos\frac{a'}{2},
 -a'\sin\frac{a'}{2}, 0\right), \\
 {\mathbf{a}'_2}^{\mathrm{T}} &= \left(a'\cos\frac{a'}{2},
 a'\sin\frac{a'}{2}, 0\right), \\
 {\mathbf{a}'_3}^{\mathrm{T}} &=
 \left(a'\frac{\cos\alpha'}{\cos\frac{\alpha'}{2}},0, a'\sqrt{1-\frac{\cos^{2}\alpha'}{\cos^{2}\frac{\alpha'}{2}}}\right).
\end{align*}
This means that only three cases are necessary in a code to output the
``standard'' conventional basis vectors in Cartesian coordinates, and
there is no need to provide individual cases for all 14 Bravais
lattices. Reduction of cases is available for the ``standard'' primitive
cell as outlined in the following section.

\subsection{''Standard'' primitive cell}

The transformation matrix $\bvec{P}$ converting ``standard'' conventional to
``standard'' primitive cells is defined as
\begin{align*}
 (\mathbf{a}'_{\mathrm{P}1},
 \mathbf{a}'_{\mathrm{P}2}, \mathbf{a}'_{\mathrm{P}3}) = (\mathbf{a}'_1,
 \mathbf{a}'_2, \mathbf{a}'_3)\bvec{P}.
\end{align*}

The ``standard'' primitive cell is the same as the ``standard''
conventional system for Bravais lattices without centring, and therefore
$\bvec{P}$ is the identity matrix.

\subsubsection{$mS$}
\label{sec:standard-prim-mS}

For monoclinic base-centred lattices (\textit{mS}, note the definition is
different from \textit{oS}),
\begin{align*}
 {\mathbf{a}'_{\mathrm{P}1}}^{\mathrm{T}} &= \left(\frac{a'}{2},\frac{b'}{2},0 \right),\\
 {\mathbf{a}'_{\mathrm{P}2}}^{\mathrm{T}} &= \left(-\frac{a'}{2},\frac{b'}{2},0 \right),\\
 {\mathbf{a}'_{\mathrm{P}3}}^{\mathrm{T}} &= \left(0,c'\cos\alpha', c'\sin\alpha' \right),
\end{align*}
\begin{equation}
\bvec{P} = 
   \begin{pmatrix}
    \frac{1}{2} & \frac{\bar{1}}{2} & 0\\
    \frac{1}{2} & \frac{1}{2} & 0 \\
   0 & 0 & 1
  \end{pmatrix}, \;\;\;
\bvec{P}^{-1} =
  \begin{pmatrix}
    1 & 1 & 0 \\
    \bar{1} & 1 & 0 \\
    0 & 0 & 1
  \end{pmatrix}.
\end{equation}

\subsubsection{$oS$}
\label{sec:standard-prim-oS}

For orthorhombic base-centred lattices (\textit{oS}, note the definition
is different from \textit{mS}),
\begin{align*}
 {\mathbf{a}'_{\mathrm{P}1}}^{\mathrm{T}} &= \left(\frac{a'}{2},-\frac{b'}{2},0 \right), \\
 {\mathbf{a}'_{\mathrm{P}2}}^{\mathrm{T}} &= \left(\frac{a'}{2},\frac{b'}{2},0 \right), \\
 {\mathbf{a}'_{\mathrm{P}3}}^{\mathrm{T}} &= \left(0,0,c' \right),
\end{align*}
\begin{equation}
\bvec{P} = 
   \begin{pmatrix}
    \frac{1}{2} & \frac{1}{2} & 0 \\
    \frac{\bar{1}}{2} & \frac{1}{2} & 0 \\
   0 & 0 & 1
  \end{pmatrix}, \;\;\;
\bvec{P}^{-1} =
  \begin{pmatrix}
    1 & \bar{1} & 0 \\
    1 & 1 & 0 \\
    0 & 0 & 1
  \end{pmatrix}.
\end{equation}

\subsubsection{$oI$, $tI$, and $cI$}
\label{sec:standard-prim-I}

For body-centred lattices (\textit{oI}, \textit{tI}, \textit{cI}),
\begin{align*}
 {\mathbf{a}'_{\mathrm{P}1}}^{\mathrm{T}} &= \left(-\frac{a'}{2},\frac{b'}{2},\frac{c'}{2} \right), \\
 {\mathbf{a}'_{\mathrm{P}2}}^{\mathrm{T}} &= \left(\frac{a'}{2},-\frac{b'}{2},\frac{c'}{2} \right), \\
 {\mathbf{a}'_{\mathrm{P}3}}^{\mathrm{T}} &= \left(\frac{a'}{2},\frac{b'}{2},-\frac{c'}{2} \right),
\end{align*}
\begin{equation}
\bvec{P} = 
   \begin{pmatrix}
    \frac{\bar{1}}{2} & \frac{1}{2} & \frac{1}{2} \\
   \frac{1}{2} & \frac{\bar{1}}{2} & \frac{1}{2} \\
   \frac{1}{2} & \frac{1}{2} & \frac{\bar{1}}{2}
  \end{pmatrix}, \;\;\;
\bvec{P}^{-1} =
  \begin{pmatrix}
    0 & 1 & 1 \\
    1 & 0 & 1 \\
    1 & 1 & 0
  \end{pmatrix}.
\end{equation}

\subsubsection{$oF$ and $cF$}
\label{sec:standard-prim-F}

For face-centred lattices (\textit{oF}, \textit{cF}),
\begin{align*}
{\mathbf{a}'_{\mathrm{P}1}}^{\mathrm{T}} &= \left(0,\frac{b'}{2},\frac{c'}{2} \right), \\
{\mathbf{a}'_{\mathrm{P}2}}^{\mathrm{T}} &= \left(\frac{a'}{2},0,\frac{c'}{2} \right), \\
{\mathbf{a}'_{\mathrm{P}3}}^{\mathrm{T}} &= \left(\frac{a'}{2},\frac{b'}{2},0 \right),
\end{align*}
\begin{equation}
\bvec{P} = 
   \begin{pmatrix}
   0 & \frac{1}{2} & \frac{1}{2} \\
   \frac{1}{2} & 0 & \frac{1}{2} \\
   \frac{1}{2} & \frac{1}{2} & 0
  \end{pmatrix}, \;\;\;
\bvec{P}^{-1} =
  \begin{pmatrix}
    \bar{1} & 1 & 1 \\
    1 & \bar{1} & 1 \\
    1 & 1 & \bar{1}
  \end{pmatrix}.
\end{equation}

\section{Lattice constant dependence on BZ topology}
\label{sec:lattice-constant}

The topology of the BZ depends on the lattice constants in some Bravais
lattices. The conditions that decide the topology, using
crystallographic conventional \cite{Bilbao2} and ``standard'' \cite{SC}
conventional lattice constants, are outlined below. Special cases for
``virtual'' crystals are omitted.

\subsection{$cP$, $cF$, $cI$, $tP$, $oP$, $oI$, $oS$, $hP$, and $mP$}
\label{sec:lattice-constant-P}

There is only one topology for each Bravais lattice.

\subsection{$tI$}
\label{sec:lattice-constant-tI}

There are two topologies for the body-centred tetragonal lattice as given in Table \ref{table:topology-bct}.

\begin{table*}[ht]
 \caption{\label{table:topology-bct}BZ topology of a body-centred tetragonal lattice.}
\centering
 \begin{tabular}{ccc}
\hline 
\hline 
Crystallography & ``Standard'' & Topology \\
\hline 
  $c > a$ & $c' > a'$ &  
 Truncated octahedron \\
  $c < a$ & $c' < a'$ &  
 Elongated dodecahedron \\
   \hline 
\hline
 \end{tabular}
\end{table*}

\subsection{$oF$}
\label{sec:lattice-constant-oF}

There are two topologies for the face-centred orthorhombic lattice. The
condition for the crystallographic conventional cell depends on the
order of basis vector lengths and is rather involved. The ``standard''
conventional cell relabels the crystallographic conventional basis
vectors such that $a'<b'<c'$, and the condition for each topology can be
written easily using ``standard'' conventional basis vector lengths as
given in Table \ref{table:topology-fco}. Then the
condition for the crystallographic conventional cell is obtained using
that for the ``standard'' conventional cell and
Table~\ref{table:symbol-ort}.

\begin{table*}[ht]
 \caption{\label{table:topology-fco}BZ topology of a face-centred orthorhombic lattice.}
\centering
 \begin{tabular}{ccc}
\hline 
\hline 
Crystallography & ``Standard'' & Topology \\
\hline 
  See the text in Sec.~\ref{sec:lattice-constant-oF} & $\frac{1}{a'^{2}}<\frac{1}{b'^{2}}+\frac{1}{c'^{2}}$ &  
 Truncated octahedron \\
  See the text in Sec.~\ref{sec:lattice-constant-oF} & $\frac{1}{a'^{2}}>\frac{1}{b'^{2}}+\frac{1}{c'^{2}}$ &  
 Elongated dodecahedron \\
   \hline 
\hline
 \end{tabular}
\end{table*}

\subsection{$hR$}
\label{sec:lattice-constant-hR}

There are two topologies for the rhombohedral lattice as given in Table \ref{table:topology-rhombo}.

\begin{table*}[ht]
 \caption{\label{table:topology-rhombo}BZ topology of a rhombohedral lattice.}
\centering
 \begin{tabular}{ccc}
\hline 
\hline 
Crystallography & ``standard'' & Topology \\
\hline 
  $\sqrt{3}a<\sqrt{2}c$ & $\alpha'<90^{\circ}$ &  
 Truncated octahedron \\
  $\sqrt{3}a>\sqrt{2}c$ & $\alpha'>90^{\circ}$ &  
 Rhombic dodecahedron \\
   \hline 
\hline
 \end{tabular}
\end{table*}

\subsection{$mS$}
\label{sec:lattice-constant-mS}

There are three topologies for the base-centred monoclinic lattice as
given in Table \ref{table:topology-monocli}. The crystallographic
conventional cell is $C$-centred with unique axis $b$ and
$\beta > 90^\circ$, while a ``standard'' conventional cell is
$C$-centred with unique axis $a$ and $\alpha' <
90^\circ$. The quantity $k'_{\mathrm{P}\gamma}$ that appears in the
conditions can be expressed using ``standard'' conventional lattice
constants as 
\begin{equation}
\cos k'_{\mathrm{P}\gamma} = \frac{a'^2 - b'^2 \sin^2
\alpha'} {a'^2 + b'^2 \sin^2 \alpha'}. \nonumber
\end{equation}
The reasoning behind how the
conditions are determined is provided in \ref{sec:appendix-b}.

\begin{table*}[ht]
 \caption{\label{table:topology-monocli}BZ topology of a monoclinic lattice.}
\centering
 \begin{tabular}{ccc}
\hline 
\hline 
Crystallography & ``Standard'' & Topology \\
\hline 
  $b<a\sin\beta$ & $k'_{\mathrm{P}\gamma}>90^{\circ}$ &  
 Truncated octahedron \\
  $\left\{ \begin{aligned} 
            &b>a\sin\beta,  \\
            &-\frac{a\cos\beta}{c}+\frac{a^{2}\sin^{2}\beta}{b^{2}}<1
  \end{aligned} \right.
      $ &
          $
          \left\{ \begin{aligned}
                  &k'_{\mathrm{P}\gamma}<90^{\circ}, \\
                  &\frac{b'\cos\alpha'}{c'}+\frac{b'^{2}\sin^{2}\alpha'}{a'^{2}}<1
                 \end{aligned} \right.
                  $ &  
 Elongated dodecahedron \\
  $
  \left\{\begin{aligned}
          &b>a\sin\beta, \\
          &-\frac{a\cos\beta}{c}+\frac{a^{2}\sin^{2}\beta}{b^{2}}>1
          \end{aligned} \right.
$ & $
          \left\{\begin{aligned}
                  &k'_{\mathrm{P}\gamma}<90^{\circ}, \\
                  &\frac{b'\cos\alpha'}{c'}+\frac{b'^{2}\sin^{2}\alpha'}{a'^{2}}>1
                 \end{aligned} \right. 
     $ &  
Truncated octahedron \\
   \hline 
\hline
 \end{tabular}
\end{table*}

\subsection{$aP$}
\label{sec:lattice-constant-aP}

The reciprocal ``standard'' conventional cell for a triclinic lattice is
a Niggli reduced cell in reciprocal space where the axes are relabeled
such that the interaxial angles are either all larger (all-obtuse) than
or all smaller (all-acute) than $90^\circ$ and $k'_\gamma$ is the
closest to $90^\circ$. Nevertheless there is only one topology of the BZ, that is, the truncated
octahedron.  The proof of this is given in \ref{sec:appendix-c}. There is
a distinct difference between the all-obtuse and all-acute reciprocal
``standard'' conventional cells regarding the orientation of reciprocal
basis vectors in the BZ. All reciprocal basis vectors
penetrate hexagonal faces in an
all-obtuse reciprocal ``standard'' conventional
cells. Our investigation of BZs of
all-acute triclinic structures using the Materials Project
database ~\cite{MatProj}
shows discrepancies from the claims that Setyawan and Curtarolo
\cite{SC} make. Setyawan and Curtarolo writes that the BZ is a truncated
octahedron and the faces of the BZ where ``standard'' reciprocal basis
vectors $\mathbf{a}'^*_1$, $\mathbf{a}'^*_2$, and $\mathbf{a}'^*_3$
penetrate are
hexagonal, hexagonal, and quadrilateral (parallelogram),
respectively. We find that $\mathbf{a}'^*_1$
penetrates a parallelogram
face and $\mathbf{a}'^*_2$ and $\mathbf{a}'^*_3$
 penetrates hexagonal faces in
the structure of
Ag$_2$S, $k'_a=0.221$ \AA$^{-1}$, $k'_b=0.229$ \AA$^{-1}$̊, $k'_c=0.105$
\AA$^{-1}̊$, $k'_\alpha=80.80^\circ$, $k'_\beta=79.36^\circ$, and
$k'_\gamma=83.58^\circ$.  On the other hand,
$\mathbf{a}'^*_2$ penetrates
a parallelogram face and $\mathbf{a}'^*_1$ and
$\mathbf{a}'^*_3$ penetrates hexagonal
faces in
the structure of H$_3$O$_5$CuSe,
$k'_a=0.184$ \AA$^{-1}̊$, $k'_b=0.207$ \AA$^{-1}̊$, $k'_c=0.141$
\AA$^{-1}̊$, $k'_\alpha=71.00^\circ$, $k'_\beta=75.21^\circ$, and
$k'_\gamma=77.19^\circ$, while $\mathbf{a}'^*_3$
penetrates a parallelogram
face and $\mathbf{a}'^*_1$ and $\mathbf{a}'^*_2$
penetrates  hexagonal faces in
the structure of B$_3$O$_7$Rb$_2$,
$k'_a=0.095$ \AA$^{-1}$, $k'_b=0.102$ \AA$^{-1}$, $k'_c=0.156$
\AA$^{-1}$, $k'_\alpha=78.58^\circ$, $k'_\beta=76.56^\circ$, and
$k'_\gamma=88.15^\circ$.
In short, always one of $\{\mathbf{a}'^*_1, \mathbf{a}'^*_2,
\mathbf{a}'^*_3\}$ penetrates a
parallelogram face and two penetrates hexagonal faces when the ``standard'' cell is
all-acute, but which one
penetrates a parallelogram depends on the lattice
constants. In light of this complexity, it is
recommended to use the parallelepiped of reciprocal basis vectors
\cite{Bilbao2,CDML} to describe band diagrams of triclinic structures unless there is a
very strong reason not to do so.

\section{Issues regarding special $\mathbf{k}$-vector point definition}
\label{sec:issues-regarding}

The philosophy behind labeling of reciprocal space is
very different between the database on the Bilbao Crystallographic
Server \cite{Bilbao2,Bilbao,CDML} and Setyawan and Curtarolo
\cite{SC}. In short, the Bilbao Crystallographic Server defines and uses a set of labels for each
type of representation domain, which is described afterwards, and BZ topology. However, fractional
coordinates of irrational special $\mathbf{k}$-vector points are not
provided. On the other hand, Setyawan and Curtarolo provides a
set of special $\mathbf{k}$-vector point symbols, including irrational special $\mathbf{k}$-vector points, for every BZ topology
in each Bravais lattice. The drawbacks are that their ``standard''
conventional basis vectors are different from the crystallographic
conventional basis vectors in many cases and that their definitions are not based on the
representation domain.

A number of concepts have to be introduced before looking into how labels of reciprocal space are defined in the Bilbao Crystallographic Server \cite{Bilbao2,Bilbao,CDML}.
The \textit{reciprocal lattice} is represented using basis vectors in reciprocal space, $\{\mathbf{a}^*_1,
      \mathbf{a}^*_2, 
      \mathbf{a}^*_3\}$. The \textit{reciprocal space group}, which is defined as a semidirect product of the point group and the translational group of the reciprocal lattice, is isomorphic with the \textit{symmorphic space group}. This means that discussion on a reciprocal space group type can be carried out using the corresponding symmorphic space group type in direct space. As a result, concepts defined in direct space, for instance Wyckoff positions and the asymmetric unit that are defined in the ITA \cite{ITA}, can be applied to the reciprocal space group. A \textit{representation domain}, which is a simply connected parts of the BZ
that contains exactly one reciprocal space vector ($\mathbf{k}$-vector) of each orbit of
$\mathbf{k}$, is assigned to each reciprocal space group type. The definitions of all 73 representation domain types are available on the Bilbao Crystallographic Server. Orbits of $\mathbf{k}$ in reciprocal space can be
categorized into symmetry-equivalent $\mathbf{k}$ vectors that
correspond to point orbits, or Wyckoff positions, of the relevant direct space
symmorphic space group \cite{Bilbao2}.

Labels of reciprocal space for every representation domain type 
are defined in the Bilbao Crystallographic Server \cite{Bilbao2,Bilbao,CDML} using Wyckoff positions of the corresponding direct space symmorphic space group type. Wyckoff positions can have zero, one, two, or three independent
coordinate variables; therefore, labels are defined for special
$\mathbf{k}$ vector points, lines (line segments, in practice), planes,
or general positions (GP), respectively. The number of labels per Wyckoff
position is not necessarily one. For instance, coordinates of Wyckoff
position $1a$ (point symmetry $mm2$) of space group type $Pmm2$ (No.
25) can be written as $(0,0,z)$ using coordinates based on the ITA
description. Labels defined for this Wyckoff position in the Bilbao
Crystallographic Server are $\Gamma$($GM$) ($z=0$), $Z$ ($z=1/2$), $\Lambda$($LD$) ($0<z<1/2$), and
$LE$ ($-1/2<z<0$) \cite{Bilbao2,Bilbao,CDML}. The names of labels depend
on the BZ topology in body-centred tetragonal, face-centred
orthorhombic, and rhombohedral Bravais lattices and the relation between
basis vector lengths in base-, body-, and face-centred orthorhombic
Bravais lattices.  The topology of the BZ and relative orientation of
basis vectors are not explicitly considered in monoclinic, base-centred
monoclinic, and triclinic Bravais lattices, and therefore only one set
of labels are defined per representation domain.

Band diagrams are drawn along a path in reciprocal space. The path is typically along line segments
connecting special $\mathbf{k}$-vector points on the BZ
surface as well as line segments connecting the
$\Gamma$ point and special $\mathbf{k}$-vector
points. As a result, symbols of special $\mathbf{k}$-vector points
on the BZ surface are the only relevant labels of interest besides the $\Gamma$ point in band
diagram calculations. In addition to labels, information is also necessary on fractional coordinates of special $\mathbf{k}$-vector points where the band diagram path bends.

The fractional coordinates of centers of faces are
always rational in a ``standard'' cell because the $\mathbf{k}$-vector of a face center has the form
\begin{equation}
\frac{p}{2}\mathbf{a}'^*_{\mathrm{P}1}+
\frac{q}{2}\mathbf{a}'^*_{\mathrm{P}2}+
\frac{r}{2}\mathbf{a}'^*_{\mathrm{P}3}
\;\;\text{where}\;\; p, q, r \in \{-1, 0,1\}. \nonumber
\end{equation} 
In contrast, some BZ vertices, and therefore center of edges, are
irrational special $\mathbf{k}$-vector points in a number of Bravais
lattices (see the definition of irrational special
$\mathbf{k}$-vector points at the end of
Section~\ref{sec:summary-of-definitions}). Table \ref{table:breakdown}
shows the distribution of Bravais lattices and BZ types among the 58,055
crystal structures in the Materials Project
database \cite{MatProj} (obtained May 25,
2015). We find that crystals with
irrational BZ vertices account for 59.8\% of the structures in
 this database. Still
46.6\% remains even if we choose to not consider triclinic cells.

One might argue that forcing usage of the first BZ is not always a good
idea when drawing a band diagram because line segments connecting
special $\mathbf{k}$-vector points are not necessarily on high symmetry
lines and, therefore, band paths should span the asymmetric unit that is uniquely defined for each representative domain.  However, taking this viewpoint does not mean that defining
symbols and identifying fractional coordinates of irrational special
$\mathbf{k}$-vector points is unnecessary. For example, consider space
group type $I4$ (No. 79) with $c/a<1$ (elongated dodecahedron BZ). The
definitions of labels in the Bilbao Crystallographic Server
\cite{Bilbao2,Bilbao,CDML} and coordinates according to the ITA
description is used in this paragraph. Line segment $\Lambda$($LD$) consists of
points with coordinates $(0,0,z)$ where $0<z\leq ld_0$. The picture of
the BZ on the Bilbao Crystallographic Server shows that line segment $\Lambda$($LD$)
starts at the $\Gamma$ ($GM$) point and ends at point $Z$, and points with
coordinates $(0,0,z)$ with $ld_0 \leq z \leq 1/2$ form line segment
$VA_1$ that connects points $Z$ and $M_{0}$. The quantity $ld_0$ is very
important. Line segments $\Lambda$($LD$) and $VA_{1}$ together form a straight line
segment in reciprocal space and come in touch at point $Z$ at coordinate
$(0,0,ld_0)$. However, the two line segments differ significantly in
character because line segment $\Lambda$($LD$) connects the $\Gamma$ ($GM$) point and point
$Z$ at the surface of the BZ while line segment $VA_1$ is entirely on a
BZ edge. The quantity $ld_0$ and the definitions of points $Z$ and $M_0$
as well as line segment $VA_1$ are not provided in tables on the
Bilbao Crystallographic Server.  Although we
can deduce using Setyawan and Curtarolo \cite{SC} that
\begin{equation}
ld_0 = \frac{1+c^2}{4a^2}, \nonumber
\end{equation}
unfortunately, cumbersome conversion of basis vectors and coordinates
are necessary between the Bilbao Crystallographic Server and Setyawan and Curtarolo. The
coordinates of special $\mathbf{k}$-vector points in crystallographic
conventional cells may be derived from those of ``standard'' primitive
cells using transformation matrices. Namely, if ``standard'' primitive
basis vectors are related to those of crystallographic conventional
cells through
\begin{equation}
(\mathbf{a}'_{\mathrm{P}1},\mathbf{a}'_{\mathrm{P}2}, \mathbf{a}'_{\mathrm{P}3}) =
(\mathbf{a}'_1,\mathbf{a}'_2, \mathbf{a}'_3)\bvec{P} = 
(\mathbf{a}_1,\mathbf{a}_2, \mathbf{a}_3)\bvec{M}\bvec{P}, 
\end{equation}
the basis vectors of the reciprocal crystallographic conventional cell
are related to those of the reciprocal ``standard'' primitive cell by

\begin{equation}
   \begin{pmatrix}
    \mathbf{a}^*_{1}  \\
    \mathbf{a}^*_{2}  \\
    \mathbf{a}^*_{3}
  \end{pmatrix}
=\bvec{MP}
   \begin{pmatrix}
    \mathbf{a}'^*_{\mathrm{P}1}  \\
    \mathbf{a}'^*_{\mathrm{P}2}  \\
    \mathbf{a}'^*_{\mathrm{P}3}
  \end{pmatrix}.
\end{equation}

Naively mixing symbols in the Bilbao Crystallographic Server
\cite{Bilbao2,Bilbao,CDML} and those in Setyawan and Curtarolo \cite{SC}
is out of the question. Existing symbols attributed to the same special
$\mathbf{k}$-vector point may or may not differ based on the two
definitions; therefore, the definition must be clearly stated when
describing results. The symbol for the same point differ for example in space group type $R3m$ (No. 160) with $\sqrt{3}a<\sqrt{2}c$ (rhombic
dodecahedron BZ). Here, point $T$ in the Bilbao Crystallographic Server
 corresponds to point $Z$ in Setyawan and
Curtarolo, and the label $P$ is used to indicate a line segment
in the former whereas $P$ is a BZ edge center in the latter. Existence of
multiple change of basis matrices for some Bravais lattices gives rise
to another problematic situation. Consider the orthorhombic space group
type $Pmmn$ (No. 59). The labels of BZ vertices for the same fractional
coordinates are the same between the Bilbao Crystallographic Server
 and Setyawan and Curtarolo,
therefore the same special $\mathbf{k}$-vector point symbol represents
the same point in the two definitions as long as the crystallographic
and ``standard'' conventional cells are identical. Basis vector lengths
satisfy $a < b < c$ in VOBr  (basis
vector lengths: $a = 3.473$ \AA, $b = 3.854$ \AA, $c = 9.050$ \AA) hence
crystallographic and ``standard'' conventional cells are the same and the
special $\mathbf{k}$-vector point symbols are the same in the two
definitions. In contrast, in CdCu$_2$O$_2$
(basis vector lengths: $a = 4.002$
\AA, $b = 9.889$ \AA, $c = 3.680$ \AA) basis vectors are retaken in the
``standard'' conventional cell such that $a' < b' < c'$, therefore the
same special $\mathbf{k}$-vector point symbol specifies different points
in the BZ.

In summary, the following two issues need to be addressed when performing band diagram calculations. First, a unique choice of basis vectors must be determined for band diagram calculations. The change of basis matrix need to be readily available if the choice of basis vectors to be used in band diagram calculations is different from the crystallographic convention. Second, a suggested path together with definitions and fractional coordinates of special $\mathbf{k}$-vector points on the path should be predetermined for each representation domain type and BZ topology. Time-reversal symmetry, in which band positions at $\mathbf{k}$ becomes the same as at $\mathbf{-k}$, significantly reduces the number of representative domain types that have to be considered. Time-reversal symmetry forces inversion symmetry to exist in the representative domain type of the BZ regardless of whether the original representative domain type has inversion symmetry or not. A list of symmorphic space group types with inversion symmetry is given in Table \ref{table:inversion-symmorphic-spacegroup}. In other words, if there is time-reversal symmetry the suggested band diagram paths in reciprocal space need to be considered only for representative domain types corresponding to the 24 space group types in Table \ref{table:inversion-symmorphic-spacegroup} instead of all 73 representative domain types. 

\begin{table*}[ht]
 \caption{\label{table:breakdown}Breakdown of 58,055 crystals in the
Materials Project by Bravais lattice and BZ topology. Symbols denoted
with an asterisk ($\ast$) indicate Bravais lattices where some BZ
vertices are irrational special $\mathbf{k}$-vector points. BZ
topologies are TO: truncated octahedron, ED: elongated dodecahedron, and
RD: rhombic dodecahedron. All-obtuse and all-acute cells are
distinguished for triclinic cells. ``Other'' in base-centred monoclinic
and triclinic cells indicate that an reciprocal interaxial angle which
should be larger or smaller than $90^\circ$ is very close to $90^\circ$
and therefore the cell cannot be categorized into any case.}  \centering
 \begin{tabular}{cccccc}
\hline 
\hline 
  Crystal system & Centring & Symbol & Topology & Count & Ratio \\
  \hline 
  Triclinic & Primitive & $aP \ast$ & All & 7646 & 13.2\% \\
   &  &  & (All-obtuse) & 3831 & 6.6\% \\
   &  &  & (All-acute) & 3797 & 6.5\% \\
   &  &  & (Other) & 18 & 0.0\% \\
  Monoclinic & Primitive & $mP \ast$ &  & 8858 & 15.3\% \\ 
  & Base-centred & $mS \ast$ & All & 6919 & 11.9\% \\
   &  &  & $b<a\sin\beta$ TO & 4973 & 8.6\% \\
   &  &  & $b>a\sin\beta$ ED & 1288 & 2.2\% \\
   &  &  & $b>a\sin\beta$ TO & 641 & 1.1\% \\
   &  &  & Other & 17 & 0.0\% \\
  Orthorhombic & Primitive & $oP$ & & 7572 & 13.0\% \\
  & Base-centred & $oS \ast$ &  & 2731 & 4.7\% \\
  & Face-centred & $oF \ast$ & All & 476 & 0.8\% \\
   &  &  & ED & 338 & 0.6\% \\
   &  &  & TO & 138 & 0.2\% \\
  & Body-centred & $oI \ast$ & & 1080 & 1.9\% \\
  Tetragonal & Primitive & $tP$ & & 2696 & 4.6\% \\
  & Body-centred & $tI \ast$ & All & 3544 & 6.1\% \\
   &  &  & ED & 826 & 1.4\% \\
   &  &  & TO & 2718 & 4.7\% \\
  Hexagonal & Primitive & $hP$ &  & 5496 & 9.5\% \\
  Rhombohedral & Primitive & $hR \ast$ & All & 3435 & 5.9\% \\
   &  &  & TO & 2861 & 4.9\% \\
   &  &  & RD & 574 & 1.0\% \\
  Cubic & Primitive & $cP$ & & 2096 & 3.6\% \\
  & Face-centred & $cF$ & & 4775 & 8.2\% \\
  & Body-centred & $cI$ &  & 731 & 1.3\% \\
   \hline 
\hline
 \end{tabular}
\end{table*}

\begin{table*}[ht]
 \caption{\label{table:inversion-symmorphic-spacegroup} List of symmorphic space group types with inversion symmetry. Brackets indicate the space group number.}
 \centering
 \begin{tabular}{llccl}
  \hline 
  \hline 
  Crystal system & Centring & Symbol && Space group type \\
  \hline 
  Triclinic & Primitive & $aP$ && $P\bar{1}$ (2) \\
  Monoclinic & Primitive & $mP$ && $P2/m$ (10) \\ 
  & Base-centred & $mS$ && $C2/m$ (12) \\
  Orthorhombic & Primitive & $oP$ && $Pmmm$ (47) \\
  & Base-centred & $oS$ && $Cmmm$ (65)\\
  & Body-centred & $oI$ && $Immm$ (71) \\
  & Face-centred & $oF$ && $Fmmm$ (69) \\
  Tetragonal & Primitive & $tP$ && $P4/m$ (83), $P4/mmm$ (123) \\
  & Body-centred & $tI$ && $I4/m$ (87), $I4/mmm$ (139) \\
  Hexagonal & Primitive & $hP$  && $P\bar{3}$ (147), $P\bar{3}1m$ (162), $P\bar{3}m1$ (164), \\
  & & && $P6/m$ (175), $P6/mmm$ (191) \\
  Rhombohedral & * & $hR$ && $R\bar{3}$ (148), $R\bar{3}m$ (166) \\
  Cubic & Primitive & $cP$ && $Pm\bar{3}$ (200), $Pm\bar{3}m$ (221) \\
  & Body-centred & $cI$ && $Im\bar{3}$ (204), $Im\bar{3}m$ (229) \\
  & Face-centred & $cF$ && $Fm\bar{3}$ (202), $Fm\bar{3}m$ (225) \\
  \hline 
  \hline
  * Primitive or triple hexagonal.
 \end{tabular}
 \end{table*}

\section{Summary}
\label{sec:summary}

An algorithm to obtain ``standard'' primitive cells for efficient and
systematic band diagram calculations is derived. First, a
crystallographic conventional cell based on the definition outlined in this work is obtained, for example, by using the
spglib code \cite{spglib-project} versions 1.7.3 and higher. Next, atom positions are obtained in
Cartesian coordinates and basis vectors are retaken as necessary to
derive the ``standard'' conventional cell. Finally, the atom positions
are converted to fractional coordinates and the basis vectors of the
``standard'' primitive cell are determined in Cartesian
coordinates. Band diagrams often pass through special
$\mathbf{k}$-vector points with irrational coordinates, and we find that
there is a need for a crystallographic convention-friendly scheme to
describe such irrational special $\mathbf{k}$-vector points.

\section*{ACKNOWLEDGMENTS}

This work was supported by Scientific Research on Innovative Areas ``Nano Informatics'' (Grant No.25106005), Grant-in-Aid for Young Scientists (B) (Grant Nos. 26820283 and 26820284) from the Japan Society for the Promotion of Science (JSPS). 

\appendix
\section{Derivation of $\bvec{M}$ for $mS$}
\label{sec:appendix-a}

We provide simple proof that one change of basis matrix suffices for all
base-centred monoclinic cells. The reciprocal primitive basis vectors
of a base-centred monoclinic lattice can be taken in Cartesian
coordinates as
\begin{align*}
 \mathbf{a}''^*_{\mathrm{P}1} &= (p,q,0), \\
 \mathbf{a}''^*_{\mathrm{P}2} &= (-p,q,0), \\
 \mathbf{a}''^*_{\mathrm{P}3} &= (0,-(r+nq),s)
\end{align*}
using four positive variables $p$, $q$, $r$, and
$s$ and an integer $n$. The restriction that this
reciprocal cell is primitive uniquely determines $p$, $q$,
and $s$. The quantity $r$ can be unambiguously defined
such that $r<q$, and the cell is ``standard'', or
$\mathbf{a}'^*_{\mathrm{P}i}=\mathbf{a}''^*_{\mathrm{P}i}$, when $n=0$. The
direct space basis vectors of the corresponding primitive cell can be
expressed as
\begin{align*}
(\mathbf{a}''^*_{\mathrm{P}1},\mathbf{a}''^*_{\mathrm{P}2},\mathbf{a}''^*_{\mathrm{P}3})=\frac{1}{2}
\begin{pmatrix}
 \frac{1}{p} & -\frac{1}{p} & 0 \\
 \frac{1}{q} & \frac{1}{q} & 0 \\
 \frac{r+nq}{qs} & \frac{r+nq}{qs} & \frac{2}{s}
  \end{pmatrix}.
\end{align*}
The basis vector lengths and interaxial angles of the corresponding
conventional cell with basis vectors
$(\mathbf{a}''^*_{1},\mathbf{a}''^*_{2},\mathbf{a}''^*_{3})$ are
\begin{align*}
 a'' &= \frac{1}{p}, \\
 b'' &= \frac{\sqrt{(r+nq)^{2} +s^{2}}}{qs}, \\
 c'' &= \frac{1}{s},
\end{align*}
and
\begin{align*}
 \cos\alpha'' &= \frac{(r+nq)}{\sqrt{(r+nq)^{2} +s^{2}}}, \\
 \beta'' &= 90^{\circ}, \\
 \gamma'' &= 90^{\circ}. \\
\end{align*}
 The set of basis vector lengths is invariant when the change of basis matrix 
\begin{align*}
 \bvec{M}=\begin{pmatrix}
             0 & \bar{1} & 0 \\
             1 & 0 & 0 \\
             0 & 0 & 1
            \end{pmatrix}
\end{align*}
is used to transform this $a$-axis, $C$-centred cell to a $b$-axis
unique, $C$-centred cell with basis vectors $(\mathbf{a}'''_{1}
,\mathbf{a}'''_{2} ,\mathbf{a}'''_{3})$ through $(\mathbf{a}''_{1}
,\mathbf{a}''_{2} ,\mathbf{a}''_3)=(\mathbf{a}'''_{1} ,\mathbf{a}'''_{2}
,\mathbf{a}'''_{3})\bvec{M}$. The basis vectors are minimized in a
crystallographic conventional cell, therefore $n=0$. This is exactly the
condition for a ``standard'' conventional cell.

\section{BZ topology of $mS$}
\label{sec:appendix-b}


We present the reasoning behind how conditions to determine the topology
of the BZ of base-centred monoclinic cells are determined. This section
uses the shorthand notation $\mathbf{b}'_i =\mathbf{a}'^*_{\mathrm{P}i}$ and
discussion is based on ``standard'' primitive cells.

The reason why the shape of the first BZ of base-centred monoclinic
cells can be categorized into three types is discussed. Cases where
inequality relations become exact equalities are not considered. We rely
on the fact that the possible shapes of Voronoi cells are limited
\cite{Horvath} and the following lemma:

Take two reciprocal ``standard'' primitive basis vectors $\mathbf{b}'_{i}
$ and $\mathbf{b}'_{j} $, where $i,j\in \{1,2,3\}$ and $i \ne j$. The
two-dimensional (2D) BZ in the plane where $\mathbf{b}'_{i} $ and
$\mathbf{b}'_{j} $ lies is a hexagon and must:
\\
a) if $\mathbf{b}'_{i} \cdot \mathbf{b}'_{j} >0$, the points 
\begin{align*}
\frac{\mathbf{b}'_{i}}{2}, \frac{\mathbf{b}'_{j}}{2}, \frac{\mathbf{b}'_{i}-\mathbf{b}'_{j}}{2}
\end{align*}
and these negatives exist on centers of different sides of the 2DBZ, or 
\\
b) if $\mathbf{b}'_{i} \cdot \mathbf{b}'_{j} <0$, the points
\begin{align*}
 \frac{\mathbf{b}'_{i}}{2}, \frac{\mathbf{b}'_{j}}{2}, \frac{\mathbf{b}'_{i}+\mathbf{b}'_{j}}{2}
\end{align*}
and these negatives exist on centers of different sides of the 2DBZ.

As $\mathbf{b}'_{1} \cdot \mathbf{b}'_{3} =\mathbf{b}'_{2} \cdot \mathbf{b}'_{3} <0$ by definition,
\begin{align*}
 \frac{\mathbf{b}'_{1}}{2}, \frac{\mathbf{b}'_{2}}{2}, \frac{\mathbf{b}'_{3}}{2}, \frac{\mathbf{b}'_{1}+\mathbf{b}'_{3}}{2}, \frac{\mathbf{b}'_{2}+\mathbf{b}'_{3}}{2}
\end{align*}
and these negatives automatically become centers of faces. This means
that the BZ must have 10 or more faces, which forces the BZ to be either
an elongated dodecahedron (faces are four hexagons and eight
parallelepipeds), a rhombic dodecahedron (faces are 12 parallelepipeds),
or a truncated octahedron (faces are eight hexagons and six
parallelepipeds).
\\
1) If $\mathbf{b}'_{1} \cdot \mathbf{b}'_{2} <0$, or
$k'_{\mathrm{P}\gamma}>90^{\circ } $, then
$\frac{\mathbf{b}'_{1}-\mathbf{b}'_{2}}{2}$ and its negative will be
centers of faces. This is geometrically possible only when the BZ is a
truncated octahedron and $\frac{\mathbf{b}'_{1}}{2}$,
$\frac{\mathbf{b}'_{2}}{2}$, $\frac{\mathbf{b}'_{3}}{2}$ and these
negatives are at the center of hexagonal
faces. $\frac{\mathbf{b}'_{1}+\mathbf{b}'_{2}+\mathbf{b}'_{3}}{2}$ and
its negative are also on a hexagonal face.
\\
2) If $\mathbf{b}'_{1} \cdot \mathbf{b}'_{2} >0$, or
$k'_{\mathrm{P}\gamma}<90^{\circ } $, there are two cases. One is when
$\frac{\mathbf{b}'_{1}+\mathbf{b}'_{2}+\mathbf{b}'_{3}}{2}$ is \textit{not} on
a face and the BZ is an elongated dodecahedron, and the remaining case
is when $\frac{\mathbf{b}'_{1}+\mathbf{b}'_{2}+\mathbf{b}'_{3}}{2}$
\textit{is} on a face and the BZ is a truncated octahedron.

\section{BZ topology of $aP$}
\label{sec:appendix-c}

We show that the BZ must be a truncated octahedron if there is no pair
of basis vectors that are perpendicular
to each other. According to Horv\'ath \cite{Horvath}, the five
topologies that a BZ can take is the parallelepiped, hexagonal prism,
rhombic dodecahedron, elongated dodecahedron, and truncated
octahedron. The concept of a ``zone'' is introduced, where any edge
\textit{e} of a Brillouin zone determines a zone of faces in which each
face has two sides equal and parallel to the given edge \textit{e}. The
number of the opposite pairs of relevant faces corresponding to a given
zone is two or three, and if it is two then the corresponding faces are
orthogonal to each other \cite{Horvath}. This means that a pair of
 basis vectors of the Brillouin zone must
be perpendicular to each other if the number of pairs of faces in a zone
is two. In a triclinic cell,  basis
vectors are generally not perpendicular to each other and therefore all
zones should consist of three pairs of faces. Edges that are shared by
hexagons in an elongated dodecahedron constitute a zone of two pairs of
hexagons, thus the BZ of a triclinic cell cannot be an elongated
dodecahedron. The rhombic dodecahedron is a special case of the
elongated dodecahedron where the length of edges shared by hexagons is
0, and at least one rectangular cross-section where the centers of edges
are penetrated by basis vectors always
exist. Therefore, the rhombic dodecahedron also cannot be a BZ of a
triclinic lattice. One can easily show that zones with two pairs of
faces exist in a hexagonal prism and in a parallelepiped, hence these
also cannot be BZs of a triclinic cell. In summary, the BZ of a
triclinic cell must be a truncated dodecahedron.

\section*{References}

\bibliography{choice}

\end{document}